\documentclass[twoside,twocolumn,9pt]{article}
\usepackage{extsizes}
\usepackage[super,sort&compress,comma]{natbib} 
\usepackage[version=3]{mhchem}
\usepackage[left=1.5cm, right=1.5cm, top=1.785cm, bottom=2.0cm]{geometry}
\usepackage{balance}
\usepackage{mathptmx}
\usepackage{sectsty}
\usepackage{graphicx} 
\usepackage{lastpage}
\usepackage[format=plain,justification=justified,singlelinecheck=false,font={stretch=1.125,small,sf},labelfont=bf,labelsep=space]{caption}
\usepackage{float}
\usepackage{fancyhdr}
\usepackage{fnpos}
\usepackage[english]{babel}
\addto{\captionsenglish}{%
  }
\usepackage{array}
\usepackage{droidsans}
\usepackage{charter}
\usepackage[T1]{fontenc}
\usepackage[usenames,dvipsnames]{xcolor}
\usepackage{setspace}
\usepackage[compact]{titlesec}
\usepackage{hyperref}
\usepackage[switch]{lineno}
\usepackage{upgreek}
\usepackage{multicol}
\usepackage{bm}
\usepackage{epstopdf}

\definecolor{cream}{RGB}{222,217,201}

\begin{document}

\pagestyle{fancy}
\thispagestyle{plain}
\fancypagestyle{plain}{
\renewcommand{\headrulewidth}{0pt}
}

\makeFNbottom
\makeatletter
\renewcommand\LARGE{\@setfontsize\LARGE{15pt}{17}}
\renewcommand\Large{\@setfontsize\Large{12pt}{14}}
\renewcommand\large{\@setfontsize\large{10pt}{12}}
\renewcommand\footnotesize{\@setfontsize\footnotesize{7pt}{10}}
\makeatother
\newcommand{\red}[1]{\textcolor{red}{#1}}

\renewcommand{\thefootnote}{\fnsymbol{footnote}}
\renewcommand\footnoterule{\vspace*{1pt}%
\color{cream}\hrule width 3.5in height 0.4pt \color{black}\vspace*{5pt}} 
\setcounter{secnumdepth}{5}

\makeatletter 
\renewcommand\@biblabel[1]{#1}            
\renewcommand\@makefntext[1]%
{\noindent\makebox[0pt][r]{\@thefnmark\,}#1}
\makeatother 
\renewcommand{\figurename}{\small{Fig.}~}
\sectionfont{\sffamily\Large}
\subsectionfont{\normalsize}
\subsubsectionfont{\bf}
\setstretch{1.125} 
\setlength{\skip\footins}{0.8cm}
\setlength{\footnotesep}{0.25cm}
\setlength{\jot}{10pt}
\titlespacing*{\section}{0pt}{4pt}{4pt}
\titlespacing*{\subsection}{0pt}{15pt}{1pt}

\fancyfoot{}
\fancyfoot[RO]{\footnotesize{\sffamily{1--\pageref{LastPage} ~\textbar  \hspace{2pt}\thepage}}}
\fancyfoot[LE]{\footnotesize{\sffamily{\thepage~\textbar\hspace{3.45cm} 1--\pageref{LastPage}}}}
\fancyhead{}
\renewcommand{\headrulewidth}{0pt} 
\renewcommand{\footrulewidth}{0pt}
\setlength{\arrayrulewidth}{1pt}
\setlength{\columnsep}{6.5mm}
\setlength\bibsep{1pt}

\makeatletter 
\newlength{\figrulesep} 
\setlength{\figrulesep}{0.5\textfloatsep} 

\newcommand{\topfigrule}{\vspace*{-1pt}%
\noindent{\color{cream}\rule[-\figrulesep]{\columnwidth}{1.5pt}} }

\newcommand{\botfigrule}{\vspace*{-2pt}%
\noindent{\color{cream}\rule[\figrulesep]{\columnwidth}{1.5pt}} }

\newcommand{\dblfigrule}{\vspace*{-1pt}%
\noindent{\color{cream}\rule[-\figrulesep]{\textwidth}{1.5pt}} }

\makeatother

\twocolumn[
 \begin{@twocolumnfalse}
\vspace{1em}
\sffamily

\noindent\LARGE{\textbf{Tracer-free Contactless Acoustic Microrheometry Quantifies Viscoelastic Spectrum of Phase-separated Condensates}} \\

\noindent\large{Kichitaro Nakajima$^{a,*}$, Taichi Yoshikawa$^{a}$, Yuta Suzuki$^{a}$, Shuta Nakatani$^{a}$, Kanta Adachi$^{a}$, Nobutomo Nakamura$^{a}$, Sanae Murayama$^{b}$, Hiroki Sakuta$^{c,d}$, Naoya Yanagisawa$^{c,d}$, Nadia A. Erkamp$^{e}$, Tomas Sneideris$^{f}$, Mao Fukuyama$^{g}$, Masateru Taniguchi$^{b}$, Miho Yanagisawa$^{c,d,h}$, Hirotsugu Ogi$^{a}$, Tuomas P. J. Knowles$^{f}$} \\

\large{\textbf{Abstract}}\\
\noindent\normalsize{The rheology of phase-separated condensates plays a central role in applications spanning advanced materials design and cellular processes, yet quantitative characterization of their viscoelasticity remains challenging due to the limitations of existing microrheological methods that require tracer particles or mechanical contact. Here, we establish tracer-free and contactless acoustic microrheometry as a versatile platform for quantifying the frequency-dependent complex shear modulus of single microscale condensates over 0.01–10~Hz. Using spatiotemporally controlled acoustic radiation force generated within a micro-acoustic resonator, this method deforms condensates for creep-recovery and oscillatory viscoelastic measurements. Quantitative validation using dextran condensates in a polyethylene-glycol continuous phase successfully captures their size- and frequency-dependent mechanical responses, while application to nucleic-acid condensates reveals salt-dependent internal viscoelastic changes at single-condensate resolution. By enabling quantitative dissection of condensate mechanics without invasive probes, acoustic microrheometry provides a broadly applicable framework for investigating phase-separated condensates across materials science, soft matter physics, biology, and beyond.}

\end{@twocolumnfalse} \vspace{0.5cm}
 ]

\renewcommand*\rmdefault{bch}\normalfont\upshape
\rmfamily
\section*{}
\vspace{-0.8cm}


\footnotetext{\textit{$^{a}$~Graduate School of Engineering, The University of Osaka, Yamadaoka 2-1, Suita, Osaka 565-0871, Japan.}}
\footnotetext{\textit{$^{b}$~The Institute of Scientific and Industrial Research, Osaka University, Osaka 565-0871, Japan}}
\footnotetext{\textit{$^{c}$~Komaba Institute for Science, Graduate School of Arts and Sciences, The University of Tokyo, Komaba 3-8-1, Meguro, Tokyo 153-8902, Japan.}}
\footnotetext{\textit{$^{d}$~Center for Complex Systems Biology, Universal Biology Institute, The University of Tokyo, Komaba 3-8-1, Meguro, Tokyo 153-8902, Japan.}}
\footnotetext{\textit{$^{e}$~ Bio-Organic Chemistry, Institute for Complex Molecular Systems (ICMS), Eindhoven University of Technology, Eindhoven, 5600MB, Netherlands}}
\footnotetext{\textit{$^{f}$~ Centre for Misfolding Diseases, Yusuf Hamied Department of Chemistry, University of Cambridge, CB2 1EW Cambridge, United Kingdom.}}
\footnotetext{\textit{$^{g}$~ Institute of Multidisciplinary Research for Advanced Materials, Tohoku University, 2-1-1 Katahira, Aoba-ku, Sendai, 980-8577, Japan.}}
\footnotetext{\textit{$^{h}$~Department of Physics, Graduate School of Science, The University of Tokyo, Hongo 7-3-1, Bunkyo, Tokyo 113-0033, Japan.}}
\footnotetext{\textit{$^{*}$~Email:k.nakajima@prec.eng.osaka-u.ac.jp}}


\section{Introduction}
Liquid–liquid phase separation is a ubiquitous physicochemical process that organizes matter across chemical\cite{Overbeek1957,Gong2024}, material\cite{Deshpande2019,Fu2024}, and biological systems\cite{Brangwynne2009, Hyman2014}. By demixing a homogeneous solution into multiple coexisting phases with distinct compositions, it generates meso- to microscale liquid compartments that exhibit properties markedly different from their surroundings. Such phase-separated condensates serve as functional units in diverse contexts, ranging from the fabrication of advanced materials\cite{Harrington2024, Naz2024} to the creation of compartmentalized microreactors\cite{Cao2024,Reis2024}. In living cells, liquid–liquid phase separation underlies the formation of membrane-less organelles that dynamically regulate biochemical reactions in space and time\cite{Shin2017}, establishing it as a fundamental organizational principle in both synthetic and natural systems.

The rheological properties of phase-separated condensates critically determine their functional behavior\cite{Banani2017,Li2025}. In particular, the frequency-dependent complex shear modulus encodes the viscoelastic response of the condensed phase, governing processes that range from reaction kinetics in cellular condensates to structural stability in engineered materials and microreactors\cite{Harrington2024, Wang2025}. Importantly, aberrant transitions in the material states of biomolecular condensates have been implicated in neurodegenerative disorders\cite{Patel2015, Alberti2019, Fischer2026}, underscoring the need for quantitative mechanical characterization. Nevertheless, quantitative determination of condensate viscoelasticity remains challenging due to their small length scale and the instability of the interfaces to mechanical perturbation.

To overcome these challenges, a range of experimental strategies has been developed and applied to dissect the condensate rheology. Optical-trapping microrheology measures frequency-dependent shear moduli through both passive tracking of thermally driven tracer particles\cite{Mason1995, Alshareedah2024_2} and active forcing via manipulation of tracer particles\cite{Jawerth2018, Neus2023}. Although these approaches have advanced the field\cite{Jawerth2020, Alshareedah2021, Alshareedah2024}, they inherently rely on the incorporation of probe particles, which can perturb the internal microstructure and complicate data interpretation. Complementary contact-based techniques, including micropipette aspiration\cite{Wang2021, Shen2023} and probe-based atomic force microscopy\cite{Naghilou2025}, deform condensates through externally applied pressure or localized mechanical contact. However, uncertainties arising from surface wettability, contact resistance, and interfacial deformation can limit applicable systems and quantitative accuracy\cite{Kung2019, Li2023}. Collectively, these intrinsic constraints highlight the need for a tracer-free contactless approach capable of quantifying the frequency-dependent viscoelastic response of individual condensates with minimal perturbation. Currently, rheofluidics\cite{Milani2026}, utilizing hydrodynamic force in a microfluidic channel, is suggested as a strategy of contactless rheological analysis for small droplet systems, further highlighting the importance of a tracer-free contactless approach.

Acoustic trapping is a technique that traps small objects using the acoustic radiation force (ARF), which acts on objects that possess an acoustic contrast relative to the surrounding medium in a standing pressure field\cite{Ding2012,Bruus2012}. The trapping mechanism is inherently contactless and depends on differences in density and compressibility between the objects and the medium, enabling the manipulation of a wide range of microscale objects, including colloids, droplets, and biological cells\cite{Evander2012, Ozcelik2018}. Furthermore, recent studies have demonstrated that ARF can also trap phase-separated condensates, suggesting its potential for probing their mechanical properties\cite{Tian2016, Nakajima2024}. Acoustic manipulation of tracer particles has been utilized for viscoelastic measurements of biological samples\cite{Nguyen2021, Bergamaschi2024}. However, this approach necessarily requires embedded tracer particles and immobilization of the samples on solid substrates, which limit its applicability to freely suspended condensates.

In this study, we present a tracer-free contactless acoustic microrheometry that quantifies the viscoelastic properties of single phase-separated condensates. This method utilizes the mechanical resonance of micro-cavities containing phase-separated condensates, where acoustic excitation at resonance generates a spatially structured standing pressure field. A single condensate can be trapped and deformed by ARF without physical contact. The force amplitude is precisely tunable through the driving voltage applied to the acoustic transducer, enabling controlled modulation of mechanical stress. Iterative application of impulse-like stresses allows creep–recovery measurements to extract capillary velocity from condensate shape relaxation, while cyclic modulation enables oscillatory rheology, from which the complex shear modulus is quantified. We first apply this approach to a synthetic system of dextran (DEX) condensates in a polyethylene glycol (PEG) continuous phase, whose viscosity and interfacial tension are independently characterized\cite{Furuki2024, Kamo2025}. The results capture the frequency- and size-dependent mechanical response of the phase-separated condensates. The method is further applied to condensates composed of poly-adenylic acid (poly-rA) at varying ionic strengths, demonstrating that acoustic microrheometry can resolve viscoelastic spectrum at the level of individual condensates.

\section{Results and Discussion}
\subsection{Working principle of acoustic microrheometry}
We fabricated a silicon–glass-based acoustofluidic chip incorporating cylindrical microcavities with the diameter and height of $200$~$\mu$m and $50$~$\mu$m, respectively, along a straight microchannel with the width of $42$~$\mu$m, where each cavity functions as an actuator for acoustic microrheometry (\textbf{Fig.~1a-d}). Upon excitation of the local mechanical resonance of the single cavity at approximately $4.5$~MHz, the acoustic field with a line-shaped pressure node is generated along the channel (\textbf{Fig.~1e}). The resulting acoustic field exerts the ARF on suspended particles, driving particles with positive acoustic contrast factors toward the pressure node\cite{Bruus2012}. This mechanism has previously been exploited for the line-like alignment of microparticles\cite{Wang2023}.

In the present study, we extend this principle to trap and deform the phase-separated condensates. As a proof of concept, we first employed DEX-rich condensates dispersed in a PEG-rich continuous phase (hereafter DEX condensates). This aqueous two-phase system was selected because it exhibits Newtonian behavior and its viscosity ($\eta = 0.2$~Pa·s) and interfacial tension ($\gamma =$ 40--50~$\mu$N/m) have been previously characterized by bulk rheology\cite{Furuki2024}, making it suitable for quantitatively validating the proposed method.

The DEX condensates were introduced into the cavity, and the acoustic field was generated by exciting the cavity resonance. Upon resonance excitation, the dispersed DEX condensates migrated toward the center of the cavity due to the ARF, forming a larger condensate via fusion events of multiple condensates (\textbf{Fig.~1f}). As the acoustic amplitude increased, the initially spherical condensate deformed into an elliptical shape (\textbf{Movie~S1}). This observation demonstrates the feasibility of acoustic microrheometry through condensate trapping and deformation induced by spatially controlled ARF. As the piezoelectric transducer allows simultaneous excitation of the multiple cavities, the viscoelastic behavior of multiple condensates can be evaluated at the same time (\textbf{Fig.~1g} and \textbf{Movie~S2}). This essentially accelerates data collection for multiple condensates and facilitates parametric study on the condensates mechanics.

\subsection{Mechanical modeling of condensates deformed by ARF}
Quantitative analysis requires determining the three-dimensional shape of deformed condensates. We therefore performed confocal microscopy to reconstruct their geometry. In the observation plane, DEX condensates were deformed into an ellipse with the minor axis aligned along the direction of ARF compression ($y$-axis; \textbf{Fig.~2a}), while elongation occurred along the perpendicular to the $x$-axis due to volume conservation. Cross-sectional images further revealed extension along the $z$-axis. Because the ARFs along the $x$- and $z$-axes are negligible compared with that along the $y$-axis, it primarily serves to trap the condensates without inducing deformation. We therefore approximate the deformed condensates as oblate spheroids with two equal major axes $a$ ($x$ and $z$-axis) and one minor axis $b$ ($y$-axis), conserving the volume as $V_c=\frac{4}{3}\pi R_0^3=\frac{4}{3}\pi a^2 b$.

In theory\cite{Gorkov1961,Bruus2012}, the ARF acting on a small particle, $\mathbf{F}$, originates from the Gor'kov potential $U_G$,
\begin{align}
U_{G} &= V_{c}\left[\frac{f_{1}}{2K_{m}}\langle p^{2} \rangle - \frac{3f_{2}\rho_{m}}{4}\langle \mathbf{v}^{2} \rangle\right],\\
\mathit{\bf{F}} &= -\nabla U_{G},
\end{align}
where $f_{1} = 1-\frac{K_{m}}{K_{c}}$ and $f_{2} = \frac{2(\rho_{c}-\rho_{m})}{2\rho_{c}+\rho_{m}}$. Here, $K_{m}$ and $K_{c}$ denote the bulk moduli of the medium and the particle, respectively, and $\rho_{m}$ and $\rho_{c}$ are their mass densities. $\langle p^{2} \rangle$ and $\langle \mathbf{v}^{2} \rangle$ represent the time-averaged squared pressure and velocity fields.

We consider a standing pressure field in the microcavity of the form $p \propto \sin ky \sin \omega t$, where the cavity center is defined as $y=0$. This leads to an ARF $\mathbf{F} \propto \sin 2ky$. Since the condensate radius is much smaller than the acoustic wavelength (typically $\sim 0.1\lambda$), the force can be linearized near $y=0$ as $\mathbf{F} \propto y$. Under this approximation, the ARF is modeled as a uniaxial surface traction, $\bm{T} = -\xi y \delta_s\cdot \bm{e}_y$ (\textbf{Fig.~2b}), where $\xi$, $\delta_s$, and $\mathbf{e}_y$ denote the gradient of the surface traction, the Dirac delta function localized at the condensate surface, and the unit vector along the $y$-axis, respectively. The effective uniaxial strain of the deformed condensate is defined as
\begin{equation}
\varepsilon = \frac{R_{0}-b}{R_{0}} = 1-\alpha^{-\frac{2}{3}},
\end{equation}
where $R_{0}$ and $\alpha=a/b$ denote the equilibrium radius and the aspect ratio of the condensate, respectively. The effective stress generated by the ARF inside the condensate, $\sigma_{\mathrm{v}}$, can be derived from its geometry as 
\begin{equation}
\sigma_\mathrm{v}=\frac{3}{4}\xi R_{0}\alpha^{-\frac{2}{3}}
\end{equation}
(See \textbf{Appendix A}). In this study, compressive strain and stress are defined as positive.

\subsection{Calibration of stress on condensates arising from ARF}
The acoustic microrheometer is actuated by applying voltage to the PZT transducer. Increasing the applied voltage amplitude, $V_T$, enhances the degree of equilibrium deformation of the DEX condensate (\textbf{Fig.~2c}), indicating that the stress applied to the condensate depends on the transducer driving voltage. Here, we calibrate the relationship between the applied voltage, $V_T$, and the effective stress, $\sigma_{\mathrm{V}}$, exerted on the condensate for quantitative analysis of condensate rheology.

Experimental measurements of the applied-voltage dependence of the equilibrium strain $\varepsilon_{\infty}$, defined as the strain after deformation reaches equilibrium, reveal that $\varepsilon_{\infty}$ increases nearly in proportion to $V_T^2$ (\textbf{Fig.~2d}). This observation corroborates that condensate deformation is driven by ARF, being consistent with previous studies\cite{Barnkob2011, Bruus2012} reporting that the amplitude of the ARF generated by piezoelectric transducers is proportional to $V_T^2$.

At equilibrium, deformation of condensates composed of a viscoelastic fluid that behaves as a purely viscous liquid at long times under the application of a constant stress can be solely determined by the balance between the interfacial restoring force and the external force. Based on this idea, we analyzed the mechanics of the deformation of a spherical condensate with the interfacial tension and equilibrium radius of $\gamma$ and $R_0$ into an oblate spheroid with the aspect ratio of $\alpha_2$ (See \textbf{Appendix B}). This analysis derives the form of the surface traction gradient $\xi$ as a function of the aspect ratio of the deformed condensate as
\begin{equation}
\xi(\alpha_2) =\frac{4\gamma\left[\Phi(\alpha_2)-2\right]}{R_0^2 \alpha_2^{-\frac{2}{3}}\left(1-\alpha_2^{-\frac{2}{3}}\right)},
\end{equation}
where 
\begin{equation}
\Phi(\alpha)=\alpha^{-\frac{4}{3}}\left(\alpha^2+\frac{\mathrm{tanh^{-1}}(\sqrt{1-\alpha^{-2}})}{\sqrt{1-\alpha^{-2}}}\right).
\end{equation}
Using this equation with $\varepsilon_{\infty}=1-\alpha_2^{-2/3}$, the equilibrium relationship between $\xi$ and $\varepsilon_{\infty}$ can be described parametrically by $\alpha_2$ (dotted line in \textbf{Fig.~2e}). This relationship was also investigated numerically using a finite element method (FEM) model (triangle symbols in \textbf{Fig.~2e}). Although a slight discrepancy between the theoretical and numerical results was observed in the large-strain regime ($\varepsilon_{\infty}>0.20$), due to the linear approximation employed in the theoretical model, \textbf{Eq.~5} accurately predicts the relationship between $\xi$ and $\varepsilon_{\infty}$ over most of the measured range.

The experimentally measured dependence of $\varepsilon_\infty$ on $V_T$ was then converted into the surface traction gradient under various applied voltages, $\xi(V_T)$, using \textbf{Eq.~5}. By substituting $\xi(V_T)$ and the condensate geometry into \textbf{Eq.~4}, the stress applied to the condensate was calibrated as a function of the applied voltage (\textbf{Fig.~2f}). The calibrated stresses were on the order of $\sigma_{\mathrm{V}} = 0.01$--$0.1$~Pa. Temporal modulation of the applied voltage enabled precise control of the stress applied to the condensate with a resolution on the order of 0.01~Pa. This controllability enables rheological measurements of soft microscale phase-separated condensates, as demonstrated below. It should be noted that, because there is imperfection in the fabrication of micro-cavities, this calibration should be conducted in each cavity prior to quantitative analysis.

\subsection{Acoustic creep-recovery test of DEX condensates}
For a condensate with equilibrium radius $R_0$, the characteristic relaxation time of the deformation $\tau_{CR}$ scales as $\tau_{CR} \propto \eta R_0 / \gamma$, where the ratio $\eta/\gamma$ (inverse capillary velocity, ICV) determines the deformation timescale\cite{Brangwynne2011, Gouveia2022}. To quantify the ICV, we performed a creep–recovery test of DEX condensates by intermittent application of the ARF. The strain $\varepsilon (t)$ of the condensate was analyzed as a function of time (\textbf{Fig.~3a}).

The DEX condensates exhibited an exponential creep and recovery behavior in response to the stepwise change in the ARF field (\textbf{Fig.~3b,c} and \textbf{Movie~S3}). An exponential function was fitted to the strain during recovery to extract its relaxation time $\tau_{CR}$ (\textbf{Fig.~3d}), and $\tau_{CR}$ were obtained for condensates with various $R_0$ (\textbf{Fig.~3e}). Then, the slope of $R_0$-$\tau_{CR}$ was acquired as $\tau_{CR}/R_{0}=$~3.35$\pm$0.04~s/mm. To convert $\tau_{CR}/R_{0}$ into $\eta/\gamma$, we analyzed the relaxation behavior of the oblate spheroid condensates using the FEM method and obtained the conversion constant $A=1.2$ at the initial strain of $\varepsilon (0)\sim$0.165 in a relation $\eta/\gamma=A\cdot\tau_{CR}/R_{0}$ (\textbf{SI Text~1} and \textbf{Fig.~S1}). For the DEX condensates, the measured $\eta/\gamma=$4.02$\pm$0.05~s/mm was reasonably consistent with the values calculated from the reported values ($\eta/\gamma\sim$5.00~s/mm)~\cite{Furuki2024}. 

We also performed the measurement of the ICV of DEX condensates including 10- and 20-ng/$\mu$L $\lambda$DNA polymers (hereafter DEX-DNA condensates) to alter their fluidity. Previous study reported that the addition of the $\lambda$DNA polymers increases the viscosity while keeping the interfacial tension nearly unchanged\cite{Furuki2024}. The results articulate the increase in the $\eta/\gamma$ by the addition of $\lambda$DNA, $\eta/\gamma=$4.49$\pm$0.07~s/mm, and $\eta/\gamma=$6.14$\pm$0.12~s/mm for DEX-DNA condensates with $\lambda$DNA concentrations of 10 and 20~ng/mL, respectively, which qualitatively confirms the reliability of the acoustic creep-recovery test.

These results demonstrate acoustic microrheometry developed here as a robust platform for probing capillary relaxation dynamics in phase-separated condensates. Notably, the acoustic creep–recovery method for measuring the ICV provides two key advantages over conventional approaches. First, iterative acoustic deformation–relaxation enables repeated measurements on the same condensate, allowing robust statistical characterization of relaxation dynamics at the single-condensate level, which is fundamentally inaccessible to conventional fusion assays. Second, the method is inherently scalable: parallel implementation across multiple cavities enables high-throughput ICV measurements, enhancing experimental efficiency and data acquisition.

\subsection{Acoustic microrheometry of single DEX condensates}
Amplitude modulation on the driving voltage signal enables the application of the cyclic stress at arbitrary frequencies to single condensates in micro-acoustic cavities. Using the calibration procedure described above, the applied voltage was converted into the stress exerted on the condensate, allowing quantitative characterization of frequency-dependent rheological properties from the strain amplitude and phase delay.

An amplitude-modulated voltage was applied to the transducer, $V_T=A_C \sin(2\pi f_C t)\left[1+\frac{A_M}{A_C}\sin(2\pi f_M t)\right],$
where \(A_C\) and \(f_C\) are the amplitude and frequency of the carrier voltage for exciting the cavity at resonance (\(f_C \sim 4.5~\mathrm{MHz}\)), while \(A_M\) and \(f_M\) are the amplitude and frequency of the modulation wave that induces sinusoidal deformation of the condensates (\(f_M = 0.01\)--\(10~\mathrm{Hz}\)) (\textbf{Fig.~4a}). The modulated voltage generates an oscillatory stress $\sigma(t)$ on the condensates, $\sigma(t) = \sigma_{b} + \sigma_{o}\sin(2\pi f_M t)$, where $\sigma_{b}$ and $\sigma_{o}$ represent the bias and oscillatory stress arising from the carrier and modulation signals, respectively. This oscillatory stress induces a strain response, $\varepsilon(t) = \varepsilon_{b} + \varepsilon_{o}\sin(2\pi f_M t-\delta)$, where $\varepsilon_{b}$ and $\varepsilon_{o}$ denote the bias and oscillatory strains, respectively, and $\delta$ is the phase delay between the stress and strain (\textbf{Fig.~4a}).

\textbf{Figure~4b-e} show the applied stress and corresponding strain responses of a DEX condensate with $R_0$ = 13.9~$\mu$m at modulation frequencies of 0.01, 0.1, 1, and 10~Hz, respectively (See also \textbf{Movies~S4--S6}). At all frequencies, the condensate exhibits purely sinusoidal strain responses at the modulation frequency \(f_M\), oscillating around the bias strain (black dotted lines in \textbf{Fig.~4b-e}). At $f_M$ = 0.01 and 0.1~Hz, the strain responses are nearly identical, with negligible phase delay ($\delta \sim 0$) (\textbf{Fig.~4b,c}). At $f_M$ = 1~Hz, the strain response begins to lag behind the applied stress, while the oscillation amplitude remains nearly constant (\textbf{Fig.~4d}). At $f_M$ = 10~Hz, the phase delay becomes substantial, and the strain amplitude decreases significantly (\textbf{Fig.~4e}).

Systematic variation of $f_M$ from 0.01 to 10~Hz (\textbf{Fig.~4f}) reveals that the strain amplitude remains nearly constant below 1~Hz, but it decreases sharply at higher frequencies, while the phase delay increases with the frequency. The Lissajous plots (\textbf{Fig.~4g}) further delineate the transition from elastic- to viscous-dominated mechanical response. The dissipated energy per cycle, $W_{loss}=\pi\sigma_o\varepsilon_o\sin\delta$, peaks at $f_{peak}\sim4$~Hz (\textbf{Fig.~4h}), corresponding to a characteristic relaxation time of $\tau=(2\pi f_{peak})^{-1}\sim40$~ms, being consistent with the estimated capillary-viscous timescale ($\tau\sim\eta R_0/\gamma\sim35$~ms).

The complex shear modulus at a frequency $f$, $G^{*}(f)=G'(f)+iG''(f)$, was determined from the strain amplitude and phase delay as $G'(f)=\frac{\sigma_o}{\varepsilon_o(f)}\cos\delta(f)$ and $G''(f)=\frac{\sigma_o}{\varepsilon_o(f)}\sin\delta(f)$. $G'(f)$ of the DEX condensate with $R_0 = 13.9~\mu$m exhibits a constant value (\textbf{Fig.~4i}), with $G' = 7.3 \pm 0.3$~Pa (mean $\pm$ SD over 0.01--10~Hz). Since the internal fluid of the DEX condensate is Newtonian\cite{Furuki2024} and therefore lacks an intrinsic elastic component, the measured $G'(f)$ solely reflects the apparent stiffness arising from interfacial tension (hereafter denoted as $G_{\gamma}$). The apparent stiffness can be theoretically derived as
\begin{align}
G_{\gamma} &= \frac{R_0^3}{5\gamma} \frac{\alpha_2^{-\frac{4}{3}}\xi_2^2 - \alpha_1^{-\frac{4}{3}}\xi_1^2}{\Phi(\alpha_2)-\Phi(\alpha_1)}
\end{align}
(see \textbf{Appendix C}). Using \textbf{Eqs.~3, 4, 6, and 7}, together with experimentally measured parameters at $f = 0.01$~Hz ($R_0 = 13.9~\mu$m, $\varepsilon_1 = 7.0\times10^{-2}$, $\varepsilon_2 = 7.8\times10^{-2}$, $\sigma_1 = 0.36$~Pa, and $\sigma_2 = 0.43$~Pa) and the reported interfacial tension ($\gamma = 40~\mu$N/m), the theoretical estimate yields $G_{\gamma} = 7.7$~Pa, in good agreement with the experimentally measured value ($G' = 7.3 \pm 0.3$~Pa). FEM simulations using the same parameters further corroborate the accuracy of the experimental measurements (red dotted line in \textbf{Fig.~4i}).

In contrast, $G''(f)$ of the DEX condensate varies with frequency as $G''(f)\sim Bf^{1.0}$, where $B=2.58$ was obtained by fitting the experimental data with a power function. For a Newtonian fluid, the loss modulus is simply estimated as $G''(f)=2\pi \eta f$, corresponding to $1.26f$~Pa for $\eta=0.2$~Pa$\cdot$s. Although the experimentally obtained proportional coefficient deviates from this simple estimate, the FEM simulations (blue dotted line in \textbf{Fig.~4i}) reproduce the experimental data remarkably well. This discrepancy suggests that, under oscillatory deformation of elliptical condensates, the effective proportional coefficient is modified from the simple form due to the complex internal flow field within the condensate. Because the FEM model incorporates these hydrodynamic effects, the strong agreement between experiment and simulation supports the accuracy of the measured frequency-dependent $G''(f)$.

Changes in condensate size are expected to alter $G_{\gamma}$ while leaving $G''(f)$ unchanged, because the former depends on interfacial geometry, whereas the latter depends solely on the viscosity of the internal fluid and is therefore independent of condensate size. The viscoelastic spectra of DEX condensates with $R_0 = 8.4~\mu$m and $20.3~\mu$m (\textbf{Fig.~4j,k}) capture this expected size-dependent mechanical response. The apparent stiffness is $G_{\gamma} = 13.8 \pm 0.5$, $7.3 \pm 0.3$, and $5.9 \pm 0.4$~Pa for $R_0 = 8.4$, $13.9$, and $20.3~\mu$m, respectively, which is in good agreement with the theoretical estimates based on \textbf{Eq.~7} ($G_{\gamma} = 14.2$, $7.7$, and $5.9$~Pa). These results suggest that the acoustic microrheometry successfully detected the distinct physical origins of storage and loss responses of the DEX condensates.

\subsection{Salt-dependent rheology of poly-rA condensates}
We next applied acoustic microrheometry to poly-rA condensates, whose phase behavior strongly depends on solvent ionic strength\cite{Nadia2023,Nadia2025}, to quantify their salt-dependent viscoelasticity. In the presence of 5\% PEG (MW$\sim$20,000 Da), poly-rA formed spherical condensates at NaCl concentrations of 500–1000~mM and gel-like aggregates above 1000~mM. Based on this phase behavior, rheological measurements were performed for condensates formed at 500, 600, and 700~mM NaCl.

Interfacial tension at each salt concentration is required to calibrate the stress applied to the condensate during active microrheology. The interfacial tension was then estimated from fluorescent recovery after photobleaching (FRAP, \textbf{Fig.~S2} and \textbf{Table~S1}) and creep-recovery measurements (\textbf{Fig.~5a}), which quantify effective viscosity $\eta$ and ICV $\eta/\gamma$, respectively. The measured viscosity, ICV, and estimated interfacial tension values are summarized in \textbf{Table~1}, and the values were used for the calibration of the ARF stress.

The quantified viscoelastic spectra at different salt concentrations (\textbf{Fig.~5b-d}) exhibit similar overall frequency dependence: $G'$ remains nearly constant at low frequencies and gradually increases at higher frequencies, while $G''$ generally increases across the measured frequency range and begins to approach saturation at higher frequencies, particularly at 600 and 700~mM, although the limited accessible frequency window prevents full resolution of this plateau behavior. Since the increase in $G'$ was not observed in Newtonian DEX condensates (\textbf{Fig.~4i--k}), this behavior likely reflects the intrinsic network elasticity of the internal poly-rA fluid. By modeling the internal poly-rA phase as a Maxwell fluid with elasticity $G_{\mathrm{int}}$ and viscosity $\eta$, the condensate mechanics can be described as a Maxwell viscoelastic interior acting in parallel with an interfacial-tension-derived stiffness $G_{\gamma}$, resulting in standard-linear-solid-like behavior with finite low-frequency elasticity (inset of \textbf{Fig.~5d}).

Decomposition of the whole-condensate mechanical response into interfacial elastic and internal viscoelastic contributions separately quantifies the frequency-dependent internal storage and loss moduli, $G'_{\mathrm{int}}$ and $G''_{\mathrm{int}}$ (\textbf{Fig.~5e,f}). Within the experimentally accessible frequency range, the internal storage modulus exceeds $\sim$1~Pa near 1~Hz and increases with frequency. Higher NaCl concentrations lead to greater $G'_{\mathrm{int}}$, particularly at 10~Hz where the modulus varies by approximately one order of magnitude across salt conditions. These observations suggest that elevated salt concentrations promote formation of a denser polymer network, likely through electrostatic screening of negatively charged poly-rA backbones by recruited counterions within the condensates.

The internal loss modulus, $G''_{\mathrm{int}}$, also increases in a salt concentration-dependent manner across the entire measured frequency range (\textbf{Fig.~5f}), with its relative magnitude remaining consistent with the salt-dependent molecular mobility observed by FRAP measurements (\textbf{Fig.~S2}). This result indicates that increasing NaCl concentration systematically enhances viscous dissipation by slowing internal polymer dynamics. Notably, $G''_{\mathrm{int}}$ remains greater than $G'_{\mathrm{int}}$ below 10~Hz, implying that despite increasing elastic contributions at elevated frequencies, the condensate interior remains predominantly viscous over these experimentally accessible timescales.

These results highlight the utility of acoustic microrheometry as a framework for resolving the viscoelastic behavior of phase-separated condensates by quantifying internal rheological dynamics at single-condensate resolution in a tracer-free and contactless manner. Although the currently accessible frequency range already reveals substantial salt-dependent rheological transitions in poly-rA condensates, extending measurements toward higher frequencies would enable more complete characterization of relaxation spectra, including crossover frequencies, plateau moduli, and faster network relaxation processes, further expanding the analytical power of acoustic microrheology for soft condensate systems.

\section{Conclusion}
In this study, we established acoustic microrheometry as a quantitative platform for probing the rheology of single phase-separated condensates. The designed micro-acoustic cavity generates spatiotemporally controlled ARF, allowing condensates to be trapped and deformed under precisely tunable stresses on the order of 0.01–0.1~Pa. This controllability achieves both iterative creep-recovery measurements and oscillatory active microrheology, enabling quantification of ICV and frequency-dependent complex shear moduli over 0.01–10~Hz. Measurements using DEX condensates validated the ability of this method to resolve frequency- and size-dependent mechanical responses, while application to poly-rA condensates demonstrated its capacity to dissect viscoelastic characteristics of internal fluid of the condensates and revealed salt-dependent rheological changes at single-condensate resolution. Acoustic microrheometry provides a versatile, tracer-free, and contactless approach for investigating microscale phase-separated condensate rheology, with broad potential applications across materials science, cellular biology, biophysics, and beyond.

\section*{Materials and Methods}
\subsection*{Fabrication of the acoustofludic device} A 4-inch-diameter Si (100) wafer (thickness: 500~$\mu$m; Electronics and Materials Corporation Limited) was used as the substrate. A 60~nm-thick Cr film was deposited onto the wafer surface by RF magnetron sputtering (EB1100, Canon Anelva). A positive photoresist (AZ5214E, MicroChemicals Gmb) was spin-coated onto the Cr layer at 5,500~rpm for 60~s, followed by soft baking at 100~$^{\circ}$C for 1~min. A polyester-based photomask containing the designed microchannel pattern prepared using AutoCAD (version~W.60.M.211, Autodesk Inc.) was fabricated by Takeda Tokyo Process Service Co., Ltd.

The photoresist was exposed to UV light (30~mW\,cm$^{-2}$) for 60~s through the photomask and developed in NMD-3 developer (TOKYO OHKA KOGYO CO., LTD.) for 20~s. The exposed Cr layer was etched using a Cr etchant solution (Cr-202, KANTO CHEMICAL CO., INC.) for 2~min. The remaining photoresist was removed by immersion in N,N-dimethylformamide (DMF, NACALAI TESQUE, INC.) overnight. The patterned Cr layer served as a hard mask for the subsequent dry etching process. Deep reactive ion etching was performed using an inductively coupled plasma etching system (RIE-400iPB, Samco) to achieve a target channel depth of 50~$\mu$m. After etching, the residual Cr mask was completely removed using a Cr etchant solution. The wafer was diced into nine individual chips using a dicing saw (DAD322, Disco Co., Ltd.).

Through-holes (diameter: 1~mm) were mechanically drilled at both termini of each microchannel to form inlet and outlet ports. The Si chips were sequentially rinsed with acetone and deionized water and dried at 75~$^{\circ}$C. Prior to bonding, the surfaces of the Si chips and glass plates (SD-2, Hoya Co., Ltd.) were treated in a plasma oven for 1~min. Anodic bonding was then performed at 450~$^{\circ}$C under an applied DC voltage of 800~V for 60~min. After bonding, the chip surface was rinsed with isopropanol and deionized water. A piezoelectric lead zirconate titanate (Pb(Zr,Ti)O$_3$, PZT) transducer with the fundamental resonant frequency od $\sim$4.5~MHz (Fuji Ceramics) was attached to the opposite side of the glass plate using a cyanoacrylate adhesive (CN adhesive, Tokyo Measuring Instruments Lab.).

\subsection*{Preparation of DEX condensates}
We used DEX with a molar mass of 450–650 kg/mol (DEX500k, 31392; Sigma-Aldrich, MO, USA), PEG with a molar mass of 7–9 kg/mol (PEG6k, 169–09125; FUJIFILM Wako Pure Chemical Corporation, Tokyo, Japan), FITC-labeled dextran with a molar mass of 450–650 kg/mol (FITC-DEX500k, 31392; Sigma-Aldrich, MO, USA), and $\lambda$DNA with a molecular weight of 3.15$\times$10$^{7}$~Da (48,502~Da, 318-00414, Nippon Gene). 

DEX and PEG solutions were separately dissolved in deionized water at a concentration of 20 wt\% each. After incubation for 1 day to ensure complete dissolution, the DEX and PEG stock solutions were mixed to obtain final concentrations of 6 wt\% for both components. To reach phase equilibrium, the mixture was vortexed for 1 min and subsequently centrifuged at 5,000~rpm for 1 h at room temperature. This procedure resulted in phase separation into a PEG-rich upper phase and a DEX-rich lower phase. Each phase was then carefully collected separately.

To prepare fluorescent labeled DEX-rich condensates dispersed in a PEG-rich continuous phase, the DEX-rich solution was first mixed with a 20 wt\% FITC-DEX solution at a volume ratio of 9:1 (DEX-rich:FITC-DEX). The labeled DEX-rich solution was then added to the PEG-rich phase at a volume ratio of 1:99 (DEX-rich:PEG-rich). For the preparation of DEX-DNA condensates, the $\lambda$DNA was dissolved into 10~mM HEPES buffer with pH~7.4, and then, the solution was mixed with the DEX-rich and PEG-rich solutions to be the final $\lambda$DNA concentration of 10 or 20~ng/mL.

\subsection*{Preparation of poly-rA condensates}
Lyophilized poly-rA with a chain length of 2,100–10,000 nucleotides was purchased from Roche (1010862001). The poly-rA was dissolved in nuclease-free water (AM9937, Thermo Fisher Scientific) to a concentration of approximately 10.0 g/L, as determined by absorbance measurement. The poly-rA solution was then mixed with HEPES buffer (pH 7.0), polyethylene glycol (PEG) with an average molecular weight of 20~kDa (PEG-20000, Sigma-Aldrich), and cyanine-5-labeled poly-rU (poly-rU-Cy5), which was obtained from Sigma-Aldrich via a custom oligonucleotide synthesis service. The final concentrations of poly-rA, poly-rU-Cy5, PEG, and HEPES buffer were 500~$\mu$g/mL, 10~$\mu$g/mL, 5\% (w/v), and 50~mM, respectively. The poly-rU-Cy5 consisted of 15 uracil residues conjugated to a single cyanine-5 moiety. The resulting molar concentrations of poly-rA and poly-rU-Cy5 were 0.4–1.8 $\mu$M and 2.0 $\mu$M, respectively. After thorough mixing, phase separation was induced by adding NaCl to final concentrations of 500, 600, or 700 mM.

\subsection*{Acoustic microrheometry}
The fabricated acoustic chip was mounted on the stage of an inverted microscope (Ti-2, Nikon) and secured using a custom-made plastic plate with screw holes that connected the inlet and outlet ports of the chip to the sample flow system. Microscopic observations were performed through the glass substrate of the chip using a 60$\times$ objective lens (Plan Fluor, Nikon). Fluorescence micrographs of the phase-separated droplets were acquired with an sCMOS camera (Andor Neo, Andor Technology) with a pixel size of 6.4~$\mu$m. The effective pixel size in the acquired images was approximately 107~nm.

The channel surface of the Si-glass acoustic device was rinsed with 1-vol\% alkaline detergent (Hellmanex III, Hellma GmbH \& Co.) by flowing it with a syringe pump (YSP-201, YMC) with a flow rate of 250~$\mu$L/h for 15~min. After washing them away with deionized water for 10~min with a flow rate of 250~$\mu$L/h, the surface of the channel was coated with a 1 wt\% hydrophilic non-ionic surfactant (Pluronic F-127, Sigma Aldrich) for 30~min. After flushing deionized water with a flow rate of 250~$\mu$L/h for 15~min, the device was filled with the deionized water prior to introducing the sample.

A 50-$\mu$L of the sample including phase-separated condensates were inflated into a PTFE microtube with an inner diameter of 0.35~mm (SP-19, Natsume Seisakusho) with a flow rate of 100~$\mu$L/h. Here, an overhead space in a plastic syringe was filled with an inert oil (HFE-7500, Fluorochem). The background flow was equilibrated by leaving the device for 30~min.

The voltage waveforms were generated using a dual-channel multifunction generator (WF1968, NF Corporation) and applied to the PZT transducer through a voltage amplifier with a gain of approximately 200 (200k100M, Insight Co., Ltd.). The voltage amplitude applied to the PZT transducer was typically ranged from 1 to 10~V.

For rheological analysis, the phase delay between the applied force field and the resulting deformation was measured. To this end, the generation of the voltage waveform was synchronized with the acquisition of condensate images, which contained information on the applied force and the resulting deformation, respectively. A trigger signal generated by the camera at the start of image acquisition was used to initiate voltage generation by the multifunction generator. Assuming negligible delay between voltage application to the PZT transducer and generation of the force field within the microcavity, the phase delay between the applied stress and resultant strain was quantified at each time point.

\subsection*{FRAP assay}
FRAP was performed using the confocal microscope (Nikon, AX) equipped with a 60×oil immersion objective (Nikon PLAN APO $\lambda$D). A $640$~nm laser was used to bleach a disk-shaped 1~$\upmu$m$^{2}$ area in the center of condensates. The half-lives of fluorescence recovery were obtained by analyzing FRAP kymographs using the Fiji software. We report the average and standard deviation of 5 repeats for each condition.

\subsection*{FEM analysis of condensate deformation}
A finite element method (FEM) model was implemented in COMSOL Multiphysics 6.1 (build 357; license no.~5092342) to simulate the deformation behavior of an oblate spheroidal condensate. The fluid was modeled as an incompressible Newtonian liquid with viscosity $\eta = 0.2~\mathrm{Pa\cdot s}$ and surface tension $\gamma = 4.0\times10^{-5}~\mathrm{N/m}$ under 1~atm and 293.15~K. The external surrounding medium was neglected.

The governing equations were solved using the Laminar Flow interface, based on the incompressible Navier--Stokes equations:
\begin{align}
\rho \frac{\partial \mathbf{u}}{\partial t} + \rho (\mathbf{u}\cdot\nabla)\mathbf{u} &= \nabla\cdot[-p\mathbf{I}+\mathbf{K}] + \mathbf{F},\\
\nabla\cdot\mathbf{u}&=0,
\end{align}
where the viscous stress tensor is defined as:
\begin{equation}
\mathbf{K} = \eta \left(\nabla\mathbf{u} + (\nabla\mathbf{u})^{T} \right).
\end{equation}
At the droplet interface, surface tension effects were incorporated through a weak contribution. Surface traction arising from the ARF were specified using bias and time-dependent axial traction gradients, $\xi_b$ and $\xi_o$:
\begin{equation}
T_z = -\xi_b z - \xi_o z \sin(2\pi f t),
\end{equation}
where $\xi_o = 0$ when evaluating relaxation behavior.

\subsection*{Analysis of condensate images for quantifying the strain}
The acquired fluorescence images of the condensates were analyzed to quantify the strain during deformation using ImageJ (ver.~1.54p). The images were first converted to binary format by applying the automatically determined threshold provided by the software. The contours of the binarized droplets were then fitted with elliptical curves using the \textit{Analyze Particles} function in ImageJ. To eliminate the influence of photobleaching during observation, the aspect ratio at each time point, $\alpha(t)$, was measured. The minor axis length was then calculated from $\alpha(t)$ using the equilibrium radius $R_{0}$ as $b(t)=R_{0}\alpha(t)^{-2/3}$. The strain was then calculated at each time point using \textbf{Eq.~3}

\section*{Author Contributions}
K.N., and T.P.J.K. designed and conceptualized the study. K.N., T.Y., K.A., N.N., S.M., M.T., and H.O. established experimental setups, analytical methods, and computational methods. K.N., T.Y., Y.S., S.N., K.A., H.S., and N.Y collected experimental results and analyzed the data. K.N., H.S., N.A.E., T.S. M.F., and M.Y. provided materials and methods. K.N., M.Y., H.O., and T.P.J.K. interpreted the data. K.N. wrote the original draft of the paper. All authors discussed the results, reviewed, edited and contributed to the final manuscript.

\section*{Conflicts of interest}
T.P.J.K. is a co-founder of Transition Bio. K.N., T.Y., Y.S., S.N., K.A., N.N., S.M., H.S., N.Y., N.A.E., T.S., M.F., M. T., M.Y., and H.O. declare no competing interests.

\section*{Acknowledgements}
This study was supported by the JSPS (25K00016 to K.N. and 24KK0104 to K.N. and M.F.), the JKA and its foundation (2024M-583 to K.N.), and the Dutchi National Growth Fund Big Chemistry (1420578 to N.A.E.). This work was supported by Advanced Research Infrastructure for Materials and Nanotechnology in Japan (Proposal Number:JPMXP12-25OS0023).

\bibliography{AcousticMicrorheometer.bib} 
\bibliographystyle{pnas2009} 

\renewcommand{\theequation}{A.\arabic{equation}}
\setcounter{equation}{0}
\section*{Appendix}
\subsection*{A. Effective stress of condensates by ARF}
The ARF acts on the surface of the condensate, where the acoustic contrast differs from that of the surrounding medium. Since the ARF can be described as a spatially dependent conservative force, we assume that the condensate is subjected to a surface traction, $\mathbf{T} =- \xi y \delta_s \mathbf{e}_y$, along the $y$-axis. Here, $\xi$, $\delta_s$, and $\mathbf{e}_y$ denote the gradient of the surface traction, the Dirac delta function localized at the condensate surface, and the unit vector along the $y$-axis, respectively.

To characterize the mechanical properties of the condensate in terms of stress and strain, we convert this surface traction into the effective stress, $\sigma_\mathrm{v}$, defined as
\begin{equation}
\sigma_\mathrm{v} = \frac{\int_{V_{C}} \sigma \, dV}{V_C},
\end{equation}
where $\sigma$ is the local internal stress and $V_C$ is the condensate volume.

The condensate deformed by the ARF within the microcavity can be approximated as an oblate spheroid with two major axes $a$ along the $x$- and $z$-axes and one minor axis $b$ along the $y$-axis, as confirmed by confocal observations (\textbf{Fig.~2a}).

For an infinitesimal cylindrical shell element of thickness $dx$, located at a distance $x$ from the $y$-axis, the shell height is determined by the spheroidal geometry as
\begin{equation}
y(x) = b\sqrt{1-\frac{x^2}{a^2}}.
\end{equation}
Accordingly, the local stress within this shell element and the corresponding differential volume element are expressed as
\begin{equation}
\sigma(x) = \xi y(x) = \xi b\sqrt{1-\frac{x^2}{a^2}},~~~dV = 4\pi x b\sqrt{1-\frac{x^2}{a^2}}\,dx,
\end{equation}
respectively. Substituting these expressions into Eq.~(1), the effective stress becomes
\begin{equation}
\sigma_\mathrm{v}
= \frac{4\pi b^2\xi\int_0^a x\left(1-\frac{x^2}{a^2}\right)\, dx}{V_c} = \frac{3}{4}\xi b.
\end{equation}
Introducing the equivalent spherical radius $R_0$ and the aspect ratio $\alpha$, the stress is finally given by
\begin{equation}
\sigma_\mathrm{v} = \frac{3}{4}\xi R_0 \alpha^{-\frac{2}{3}}.
\end{equation}

\subsection*{B. Relationship between the surface traction gradient and equilibrium strain}
We derive the relationship between the surface traction gradient $\xi$ and equilibrium strain $\varepsilon_{\infty}$ under the application of a constant ARF. Physical parameters in the initial and final states are described by subscripts 1 and 2, respectively.

Here, we consider the deformation of a spherical condensate ($\alpha_1=1$) with equilibrium radius $R_0$ into an oblate spheroid with aspect ratio $\alpha_2$. At equilibrium, the equilibrium strain is expressed as $\varepsilon_{\infty}=1-\alpha_2^{-\frac{2}{3}}$ from \textbf{Eq.~3} in the main manuscript. This deformation alters the system energy through change in the surface area of the condensate $A$. The change in the surface energy $\Delta E_s$ is written as $\Delta E_s = E_{s2}-E_{s1}=\gamma (A_2-A_1)$. The surface area $A$ of an oblate spheroid with major axis $a$ and minor axis $b$ is expressed as
\begin{equation}
A = 2\pi\left(a^2+\frac{b^2\tanh^{-1}(e)}{e}\right),
\end{equation}
where $e=\sqrt{1-\frac{b^2}{a^2}}$. Then, this can be expressed as a function of the aspect ratio $\alpha$ as
\begin{equation}
A(\alpha) = 2\pi R_0^{2} \alpha^{-\frac{4}{3}}\left(\alpha^2+\frac{\mathrm{tanh^{-1}}(\sqrt{1-\alpha^{-2}})}{\sqrt{1-\alpha^{-2}}}\right) = 2\pi R_0^{2}\Phi(\alpha).
\end{equation}
Here, $\Phi(\alpha)$ is defined as 
\begin{equation}
\Phi(\alpha) = \alpha^{-\frac{4}{3}}\left(\alpha^2+\frac{\mathrm{tanh^{-1}}(\sqrt{1-\alpha^{-2}})}{\sqrt{1-\alpha^{-2}}}\right).
\end{equation}
Then, the change in the surface energy is expressed as 
\begin{equation}
\Delta E_s =2\pi \gamma R_0^2\left[\Phi(\alpha_2)-\Phi(\alpha_1)\right].
\end{equation}

To correlate the gradient of the surface traction with surface tension, we equate this surface energy change to the apparent strain energy caused by the equilibrium strain $\Delta E_{\varepsilon_\infty}$ as
\begin{align}
\Delta E_{\varepsilon_\infty} &=\frac{1}{2}\sigma_{\mathrm{v}}\varepsilon_\infty V_c\\
&=\frac{\pi R_0^4 \xi \varepsilon_{\infty} \alpha_2^{-\frac{2}{3}}}{2}.
\end{align}
Here, it should be noted that the strain energy in the initial state is 0, so that the change in the strain energy can be written by parameters in the final state. By substituting $\alpha_1=1$ and $\varepsilon_{\infty}=1-\alpha_2^{-\frac{2}{3}}$ and equating \textbf{Eq.~A.9} and \textbf{A.11}, the surface traction gradient can be expressed as a function of $\alpha_2$,
\begin{equation}
\xi(\alpha_2) =\frac{4\gamma\left[\Phi(\alpha_2)-2\right]}{R_0^2 \alpha_2^{-\frac{2}{3}}\left(1-\alpha_2^{-\frac{2}{3}}\right)}.
\end{equation}
Using this expression together with $\varepsilon_{\infty}=1-\alpha_2^{-\frac{2}{3}}$, the relationship between $\xi$ and $\varepsilon_{\infty}$ at equilibrium can be described parametrically by $\alpha_2$.

\subsection*{C. Apparent stiffness arising from surface tension}

We derive the apparent stiffness arising from surface tension, denoted by $G_{\gamma}$, which represents the apparent storage modulus when the condensate is treated as an elastic body. We consider an oblate spheroidal condensate initially at equilibrium with aspect ratio $\alpha_1$ under surface traction gradient $\xi_1$, which is subsequently deformed to a new equilibrium state with aspect ratio $\alpha_2$ under an increased surface traction gradient $\xi_2$. The resulting increase in the system energy is assumed to originate solely from the corresponding increase in surface energy.

Using the surface area function $A(\alpha)=2\pi R_0^2\Phi(\alpha)$ derived above, the change in surface energy between the two equilibrium states is expressed as
\begin{equation}
\Delta E_s = 2\pi\gamma R_0^2 \left[\Phi(\alpha_2)-\Phi(\alpha_1)\right].
\end{equation}

To derive the apparent stiffness associated with this surface energy increase, we equate $\Delta E_s$ to the corresponding apparent strain energy change, $\Delta E_{\varepsilon}$. The infinitesimal strain energy stored in a cylindrical shell element subjected to local stress $\sigma(x)$ is
\begin{equation}
dE_{\varepsilon} = \frac{\sigma^2(x)}{2G_{\gamma}}\, dV.
\end{equation}
Following the derivation presented in SI Appendix A, integration over the entire spheroid yields
\begin{equation}
E_{\varepsilon}(\alpha,\xi) = \frac{2\pi R_0^5 \alpha^{-\frac{4}{3}}\xi^2}{5G_{\gamma}}.
\end{equation}
Thus, the apparent strain energy difference between the two equilibrium states is
\begin{align}
\Delta E_{\varepsilon} &= E_{\varepsilon}(\alpha_2,\xi_2) - E_{\varepsilon}(\alpha_1,\xi_1) \\
&= \frac{2\pi R_0^5}{5G_{\gamma}}\left(\alpha_2^{-\frac{4}{3}}\xi_2^2 - \alpha_1^{-\frac{4}{3}}\xi_1^2\right).
\end{align}
Using the effective stress,$\sigma_{\mathrm{v}i}=\frac{3}{4}\xi_i R_0 \alpha_i^{-\frac{2}{3}}(i=1,2)$, this relation can be rewritten as
\begin{equation}
\Delta E_{\varepsilon} = \frac{32\pi R_0^3}{45G_{\gamma}} \left(\sigma_{\mathrm{v}2}^2-\sigma_{\mathrm{v}1}^2 \right).
\end{equation}
By equating the surface energy increase and apparent strain energy increase, $\Delta E_s=\Delta E_{\varepsilon}$, the apparent stiffness due to surface tension is obtained as
\begin{align}
G_{\gamma} &= \frac{R_0^3}{5\gamma} \frac{\alpha_2^{-\frac{4}{3}}\xi_2^2 - \alpha_1^{-\frac{4}{3}}\xi_1^2}{\Phi(\alpha_2)-\Phi(\alpha_1)}\\
&=\frac{16R_0}{45\gamma}\frac{\sigma_{\mathrm{v}2}^2-\sigma_{\mathrm{v}1}^2}{\Phi(\alpha_2)-\Phi(\alpha_1)}.
\end{align}
This result indicates that the apparent stiffness induced by surface tension is governed by condensate size, surface tension, and deformation-dependent stress variations between equilibrium states.

\begin{figure*}[t]
\centering
\includegraphics[width=\textwidth]{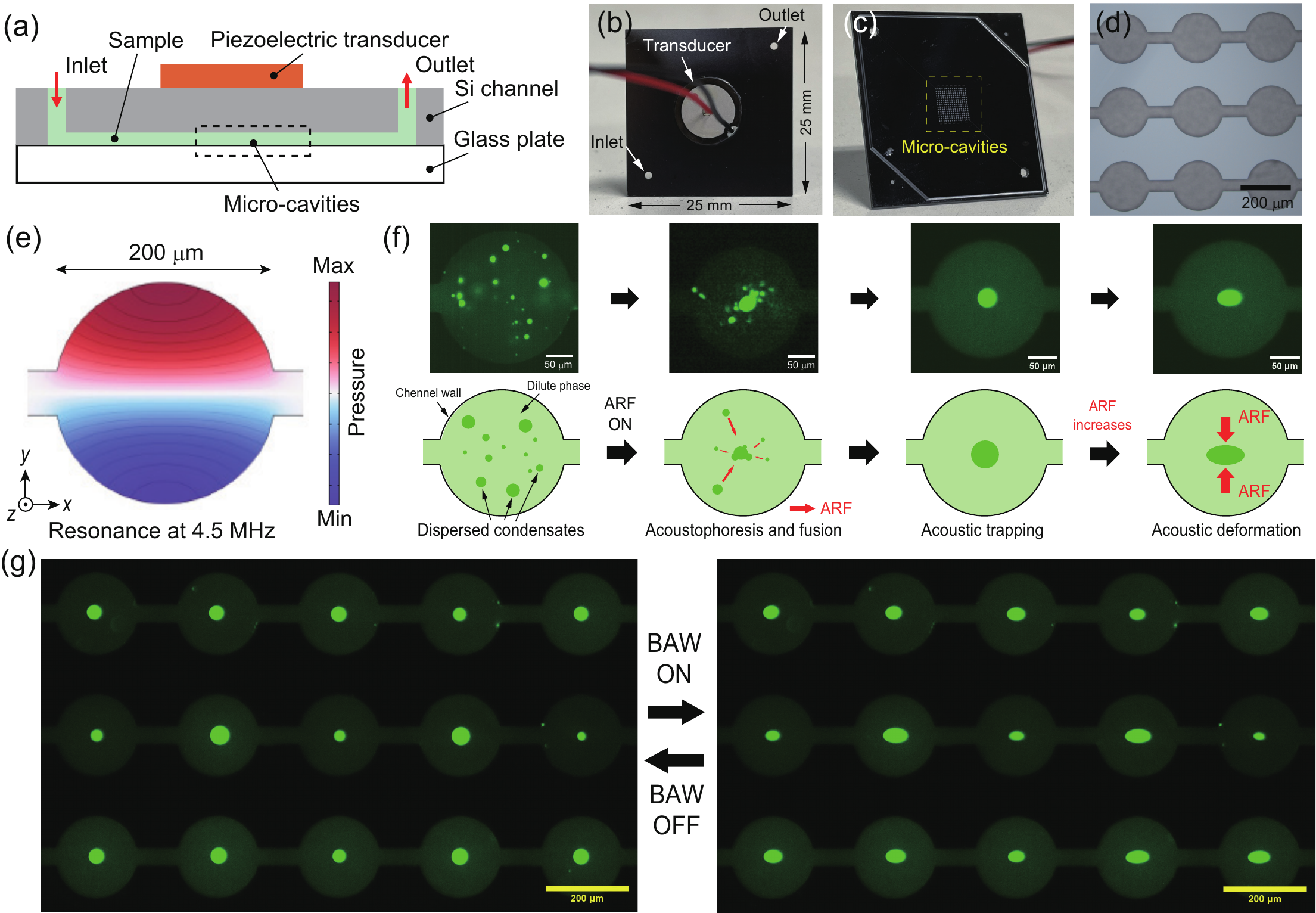}
\caption{\textbf{Contactless acoustic trapping and deformation of the DEX condensates by the acoustic radiation force generated in mechanically resonating micro-cavities fabricated on the Si-glass device.} (a) Schematic cross-sectional view of the fabricated acoustic device. (b–d) Photographs of the fabricated Si–glass device integrated with the piezoelectric transducer: (b) top view, (c) bottom view, and (d) micrograph of the microcavities located at the bottom center of the device. (e) Simulated pressure distribution within the microcavity at resonance. (f) Sequential snapshots showing acoustic trapping and deformation of a DEX condensate under the acoustic field, with a final diameter of approximately 30~$\mu$m. (See also \textbf{Movie~S1}) (g) Simultaneous trapping and deformation of multiple DEX condensates induced by ARF (See also \textbf{Movie~S2}).}
\label{fig:1}
\end{figure*}
\clearpage

\begin{figure*}[t]
\centering
\includegraphics[width=\textwidth]{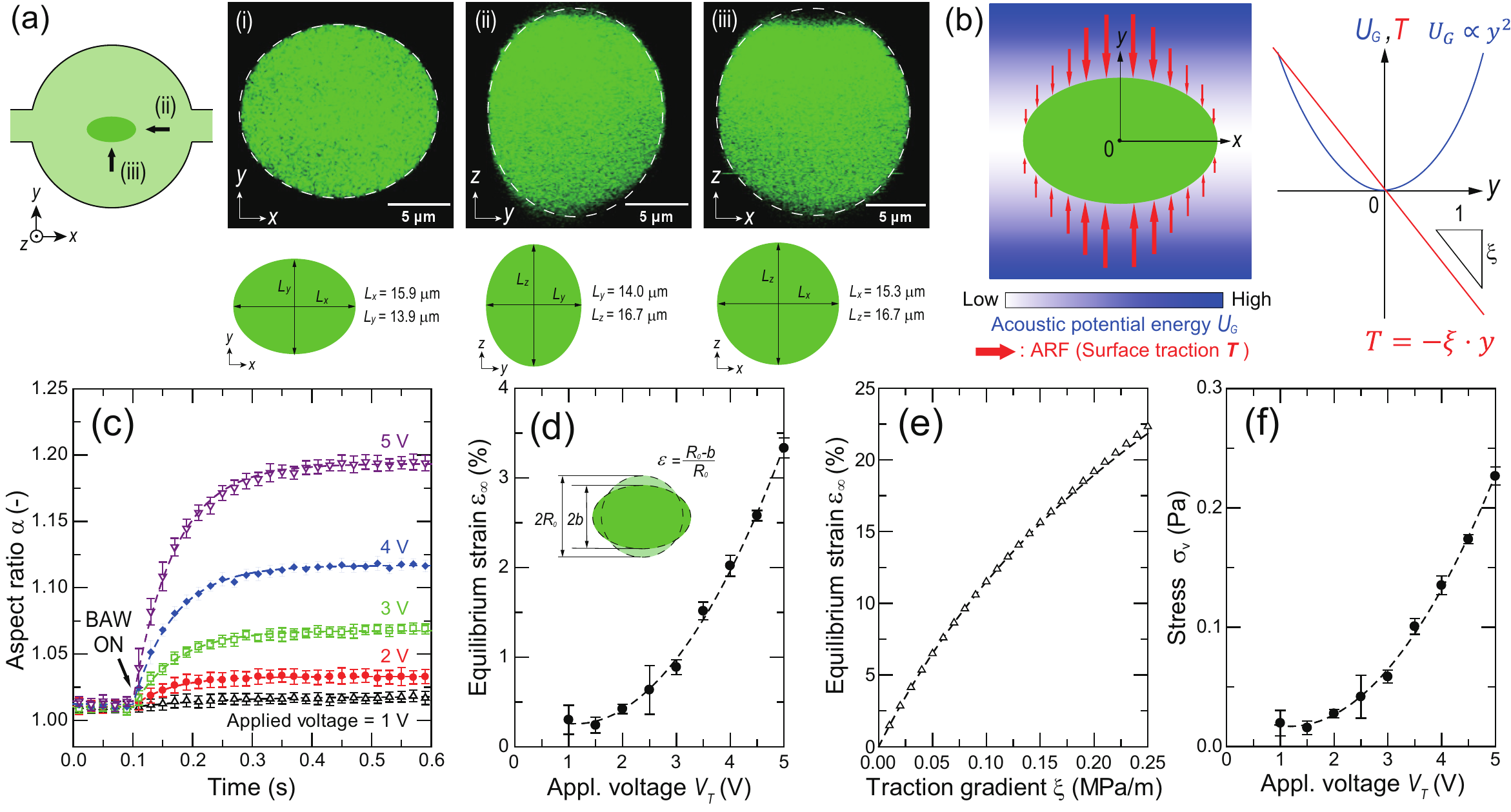}
\caption{\textbf{Mechanical modeling of condensate deformation and calibration of acoustic potential field.} (a)~Confocal 3D images of deformed DEX condensates in the (i)~$xy$-, (ii)~$yz$-, and (iii)~$xz$-planes. The definition of each axis direction is shown in the left panel. (b)~Spatial distribution of the ARF exerted on the condensate, modeled as a surface traction $\mathbf{T}$. (c)~Time-course changes in the aspect ratio $\alpha$ of DEX condensates following application of various driving voltages to the PZT transducer at 0.1~s. Solid plots represent experimentally measured values, and dashed lines indicate corresponding exponential fits of the deformation behavior used to estimate the equilibrium shape at each driving voltage. Error bars denote the standard deviation from repeated voltage applications ($n=10$). (d)~Relationship between the applied voltage and resulting equilibrium strain of a DEX condensate with a radius of 13.9~$\mu$m. Error bars represent the standard deviation from three independent measurements of the same condensate. (e)~Relationship between the surface traction gradient and resulting equilibrium strain, calculated from \textbf{Eqs.~3 and 5} (triangle plots) and the FEM model (dashed line) for a condensate with $\gamma = 40~\mu\mathrm{N/m}$ and $R_0 = 13.9~\mu\mathrm{m}$. (f)~Calibrated volume-averaged stress exerted on DEX condensates by ARF as a function of the voltage applied to the transducer.}
\label{fig:2}
\end{figure*}
\clearpage

\begin{figure*}[t]
\centering
\includegraphics[width=\textwidth]{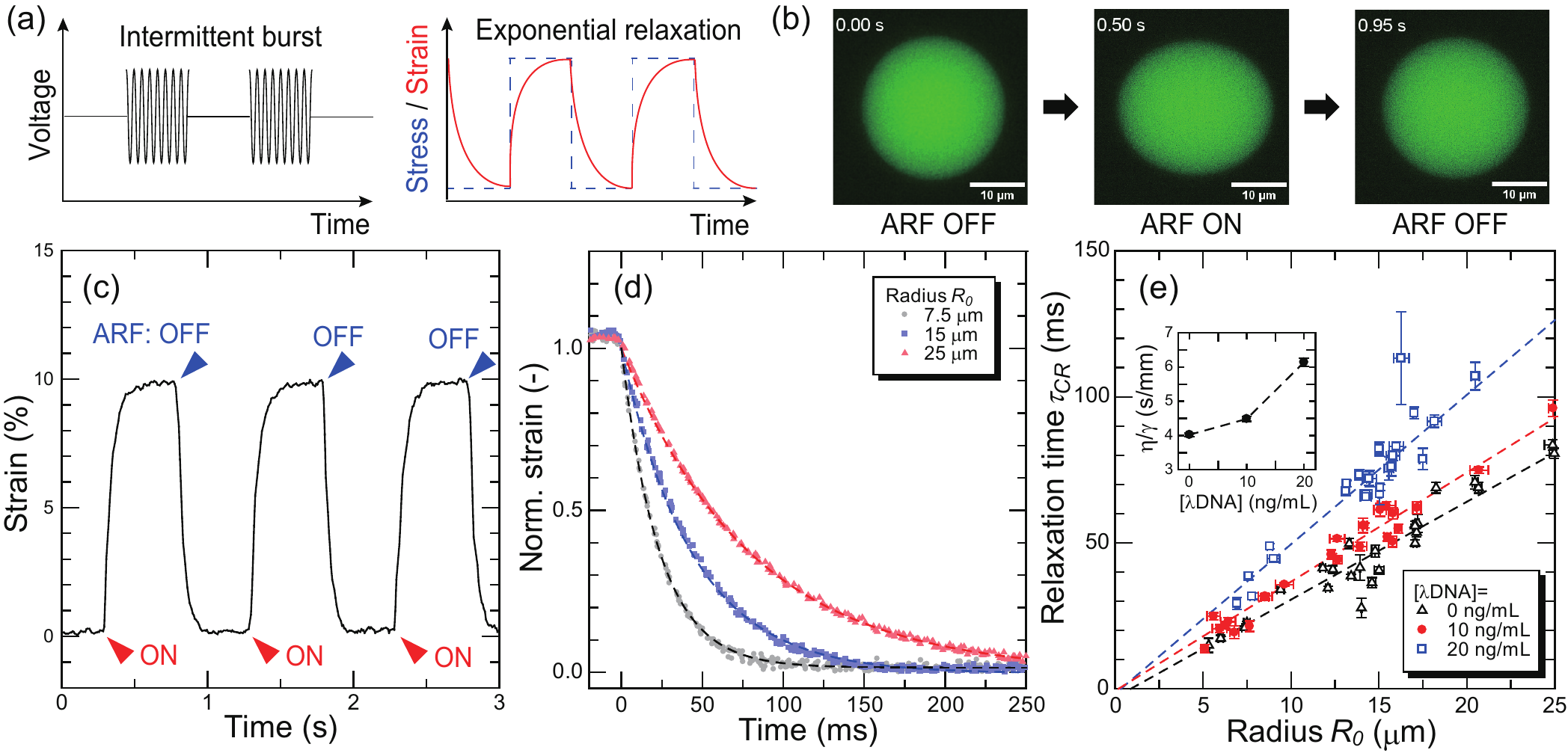}
\caption{\textbf{Creep-recovery test for analysis of inverse capillary velocity of DEX condensates.} (a)~Schematic illustration of the intermittent burst voltage driving the acoustic transducer, generated strain by the ARF, and resultant exponential relaxation of the condensate deformation. (b)~Fluorescent micrographs of the DEX condensate deformed by the ARF. The scale bars denote 10~$\mu$m. (See also \textbf{Movie~S3}) (c)~Experimental measurement of the time-course of the strain under iterative application of the ARF on the condensate shown in panel b. (d)~Exponential relaxation behavior of the deformation during the shape relaxation of the DEX condensates after removing the stress with three various radii. The plots and dashed lines represent the experimental measurements and fitted curve with the exponential function. (e)~Relaxation time of the deformation as a function of the condensate radius of the DEX condensates with and without the addition of $\lambda$DNA ($n$=25 for each condition). The dashed lines denote the fitted line of each dataset with a linear function. Inset shows the calculated inverse capillary velocity at various DNA concentrations. }
\label{fig:3}
\end{figure*}
\clearpage

\begin{figure*}[t]
\centering
\includegraphics[width=\textwidth]{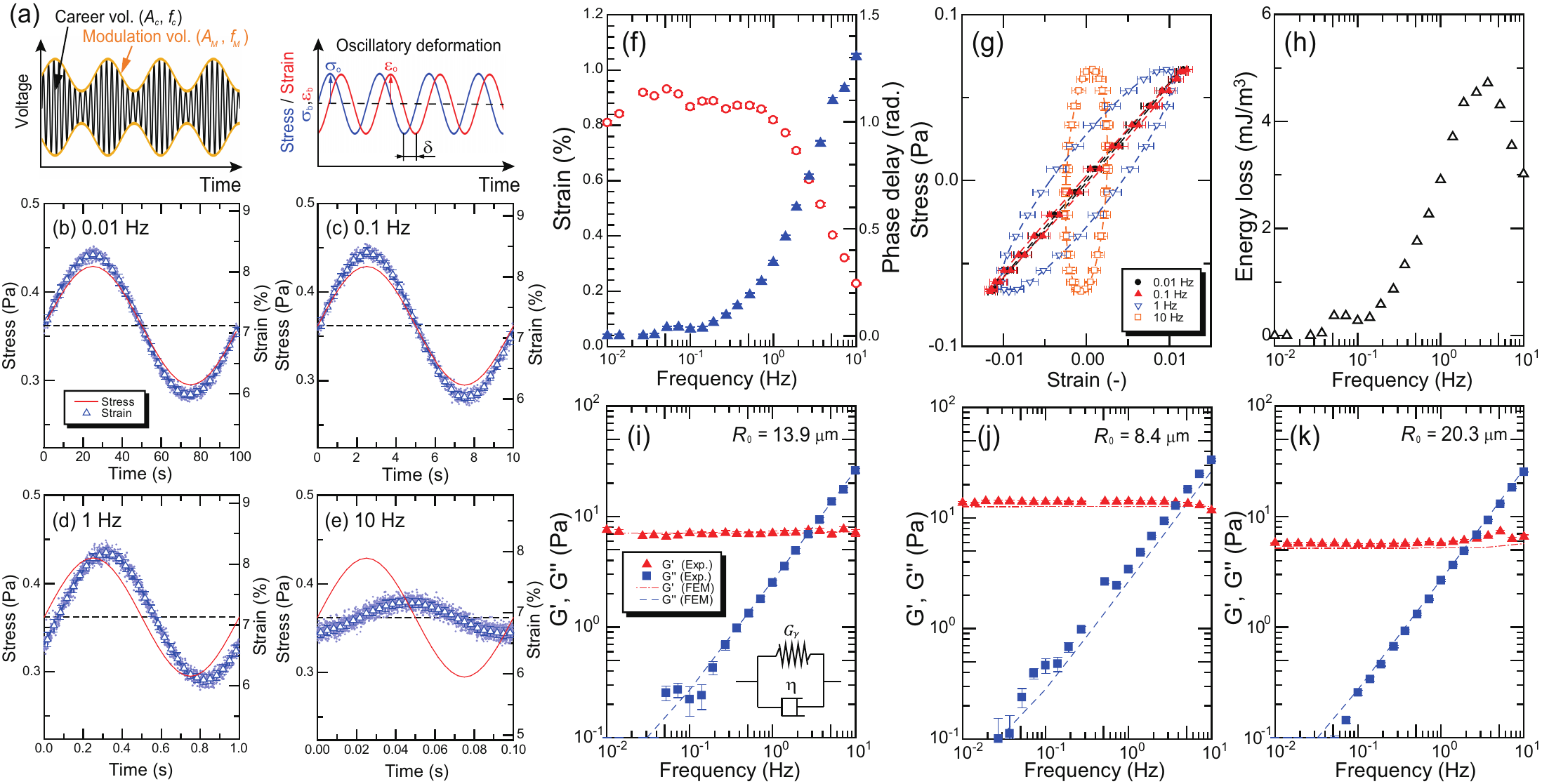}
\caption{\textbf{Active microrheology of DEX condensates.} (a) Schematic illustration of the amplitude-modulated voltage signal driving the acoustic transducer, the resulting oscillatory strain generated by ARF, and the corresponding strain response with a phase delay relative to the applied stress. (b--e) Applied oscillatory stress and resultant strain of DEX condensates with $R_0 = 13.9~\mu\mathrm{m}$ at (b)~0.01, (c)~0.1, (d)~1, and (e)~10~Hz. Small circles represent raw experimental data measured over multiple periods; hollow triangles and error bars indicate averaged data and corresponding standard deviations obtained by dividing the raw data into 30 discrete bins over one period; dashed lines represent sinusoidal fits to the averaged data with a phase delay. (f) Frequency dependence of the strain amplitude and phase delay between stress and strain over the range of 0.01--10~Hz. Error bars denote the uncertainties of the fitted parameters obtained from the covariance matrix. (g) Lissajous plots of DEX condensates with $R_0 = 13.9~\mu\mathrm{m}$ at 0.01, 0.1, 1, and 10~Hz. Dashed lines indicate fitted ellipses. (h) Energy dissipation per cycle of DEX condensates with $R_0 = 13.9~\mu\mathrm{m}$ over the frequency range of 0.01--10~Hz. (i--k) Frequency-dependent storage and loss moduli of DEX condensates with (i)~$R_0 = 13.9~\mu\mathrm{m}$, (j)~$R_0 = 8.4~\mu\mathrm{m}$, and (k)~$R_0 = 20.3~\mu\mathrm{m}$. The inset in panel~i illustrates the mechanical model describing the mechanical response of DEX condensates, consisting of an apparent elastic interfacial component, $G_{\gamma}$, and the viscosity of the internal condensate fluid, $\eta$.}
\label{fig:4}
\end{figure*}
\clearpage

\begin{figure*}[t]
\centering
\includegraphics[width=\textwidth]{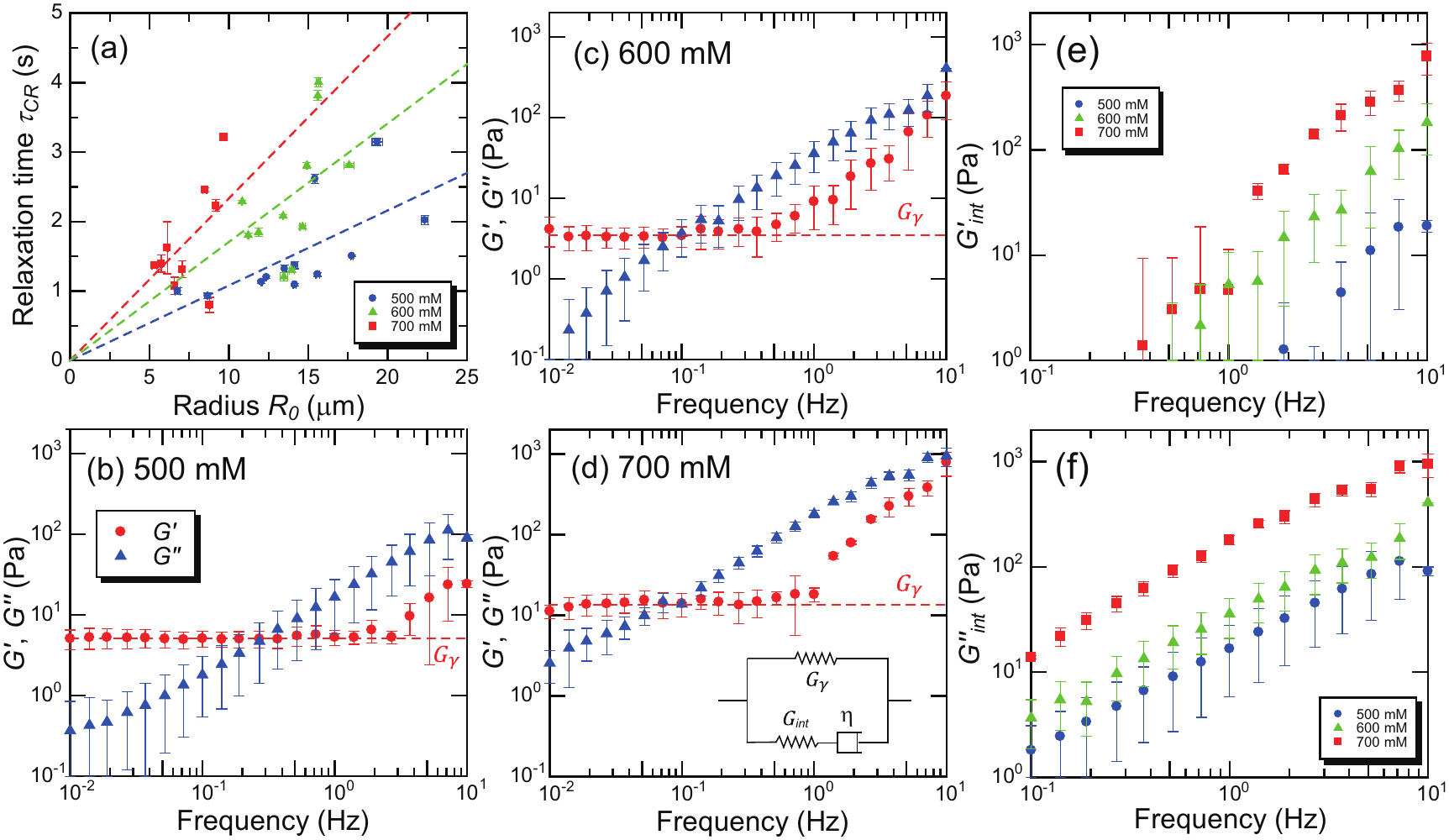}
\caption{\textbf{Salt-dependent rheological change in the poly-rA condensates.}  (a) Results of creep-recovery tests for poly-rA condensates formed at NaCl concentrations of 500, 600, and 700~mM, for analyzing the inverse capillary velocity under each condition. Error bars represent the standard deviation of the relaxation time obtained from multiple relaxation dynamics measurements ($n > 10$). (b--d) Measured viscoelastic spectra of poly-rA condensates formed with (b)~500~mM, (c)~600~mM, and (d)~700~mM NaCl. Error bars indicate the standard deviation from measurements of three independent condensates. The inset in panel~d illustrates the mechanical model of the poly-rA condensates, consisting of an apparent elastic interfacial component, $G_{\gamma}$, and a viscoelastic Maxwell fluid characterized by network elasticity ($G_{\mathrm{int}}$) and viscosity ($\eta$). (e,f) Frequency dependence of the (e)~storage modulus and (f)~loss modulus of the internal fluid of poly-rA condensates at various salt concentrations. Error bars represent the standard deviation from measurements of three independent condensates.}
\label{fig:5}
\end{figure*}

\begin{table*}[b]
\centering
\caption{Salt-concentration dependency of the physical parameters of poly-rA condensates estimated from FRAP and acoustic creep-recovery test.}
\begin{tabular}{|c|c|c|c|}
\hline
Salt concentration (mM) & Viscosity (Pa$\cdot$ s) & ICV ($\times$10$^5$ s/m) & interfacial tension ($\mu$N/m) \\
\hline
500 & 2.4$\pm$0.3 & 1.08 & 18$\pm$2 \\
\hline
600 & 3.0$\pm$0.1 & 1.71 & 15$\pm$1 \\
\hline
700 & 6.5$\pm$0.3 & 2.33 & 28$\pm$1 \\
\hline
\end{tabular}
\end{table*}
\end{document}